\documentclass[a4paper,12pt]{article}

\usepackage[latin1]{inputenc}
\usepackage{amsmath,amssymb}

\newcommand{\ie}{i.e.}
\newcommand{\eg}{e.g.}
\DeclareMathOperator{\re}{Re}
\DeclareMathOperator{\im}{Im}
\DeclareMathOperator{\ch}{ch}
\DeclareMathOperator{\Td}{Td}
\DeclareMathOperator{\Arg}{Arg}
\DeclareMathOperator{\Arc}{Arc}
\DeclareMathOperator{\Log}{Log}
\DeclareMathOperator{\Tr}{tr}
\DeclareMathOperator{\Hom}{Hom}
\DeclareMathOperator{\rk}{rk}
\newcommand{\muE}[1]{\mu_E^{(#1)}}
\newcommand{\muEp}[1]{\mu_{E'}^{(#1)}}
\newcommand{\piE}[1]{\pi_E^{(#1)}}
\newcommand{\piEp}[1]{\pi_{E'}^{(#1)}}
\newcommand{\piS}[1]{\pi_S^{(#1)}}

\begin{document}

\begin{titlepage}

\title{Non-linear Yang--Mills instantons from strings are
$\pi$-stable D-branes}

\author{H.~Enger\footnote{hakon.enger@fys.uio.no}~~and
C.~A.~L\"utken\footnote{lutken@fys.uio.no}\\
\\
\normalsize\it Theory Group, Department of Physics, University of Oslo,\\
\normalsize\it P.O.Box 1048 Blindern, NO-0316 Oslo, Norway
}

\date{ }

\maketitle

\vspace{1.5cm}

\begin{abstract}
  We show that B-type $\Pi$-stable D-branes do not in general reduce
  to the (Gieseker-) stable holomorphic vector bundles used in
  mathematics to construct moduli spaces.  We show that solutions of
  the almost Hermitian Yang--Mills equations for the non-linear
  deformations of Yang--Mills instantons that appear in the low-energy
  geometric limit of strings exist iff they are $\pi$-stable, a
  geometric large volume version of $\Pi$-stability.  This shows that
  $\pi$-stability is the correct physical stability concept.  We
  speculate that this string-canonical choice of stable objects, which
  is encoded in and derived from the central charge of the
  string-\emph{algebra}, should find applications to algebraic
  geometry where there is no canonical choice of stable
  \emph{geometrical} objects.
\end{abstract}

\thispagestyle{empty}
\end{titlepage}

\section{Introduction and summary}

Stability-ideas play a central role in both physics and mathematics,
but it is far from obvious what relationship, if any, there is between
these apparently quite distinct concepts.

Mathematicians want to classify bundles and sheaves, and find useful
compactifications of moduli spaces that can assist them in this and
other tasks.  Their starting point is the observation that all bundles
can be `derived' from the small subset of stable bundles.  A bundle
over a curve is called Mumford-($\mu$-) stable if it satisfies a
certain topological constraint.  The condition is that the ratio of
the degree to the rank of every sub-bundle $E'$ is smaller than this
ratio for the bundle $E$ itself:
\begin{align}
\mu_{E'} < \mu_E := \deg(E)/\rk(E).
\end{align}
The ratio $\mu_E$ is called the \emph{slope} of the bundle.
Equivalently, if we to each bundle associate a complex number
\begin{align}
Z(E) := i \rk(E) - \deg(E) = i \rk(E) - c_1(E)
\end{align}
this is a condition on the \emph{phase} $\varphi$ of the sub-bundles:
\begin{equation}
\varphi_{E'} < \varphi_E := \Arg Z(E) = \im \Log Z(E)
= -\Arc \cot \frac{\deg(E)}{\rk(E)}.
\end{equation}

The classification problem is now reduced to finding all the stable
bundles, whose moduli space\footnote{Some features of moduli spaces of
  $\mu$-stable bundles were reviewed by Sharpe~\cite{Sharpe:1998zu}.}
is compactified by `gluing in' the marginally stable cases, which are
called strictly semi-stable.

This is somewhat reminiscent of the physical problem of identifying
irreducible constituents from which all modes can be built, but it is
not clear how conservation laws and field equations transmute into
these topological conditions, and in fact it is not even clear from
mathematics which topological stability condition is appropriate.

In the presence of sufficient supersymmetry the bridge between the
mathematical and physical stabilities is provided by the central
extension of the superalgebra, called the central- or BPS-charge
($\mathcal{Z}$) in physics.  In the simplest low-energy, large volume
limit of string where conventional geometry is recovered, some
D-branes $\mathcal{E}$ should reduce to holomorphic bundles $E$, and
indeed one finds that ${\mathcal{Z}(E)}\rightarrow Z(E)$ in this limit
in the case where the target space is an elliptic curve.  In this
simple case $\mu$-stability can be immediately identified as
conservation of energy and charge of the brane.

While the topological interpretation of $\mathcal{Z}$ cannot be
maintained in general, the BPS-charge is a fully stringy concept, and
it is not unreasonable to expect that the condition on the phase of
this complex charge ({\ie} $\Arg\mathcal{Z}' \le \Arg\mathcal{Z}$) can
be retained as the fundamental stability
criterion~\cite{Douglas:2000ah}.  This so-called $\Pi$-stability
encodes in a simple and physically transparent manner the physics of
charge and energy conservation, and therefore suggests that this is a
fundamental concept delivered to us by string theory.

It is encouraging that $\Pi$-stability reduces to $\mu$-stability in
the one-dimen\-sional case, but it is not clear that this is the case
in general.  In one (complex) dimension $\mu$-(semi-)stability appears
to be the relevant property for classifying holomorphic bundles, but
in higher dimensions it seems that a more refined concept of
semi-stability is needed.  While there does not appear to be a
canonical way to choose this refinement, it is an empirical
observation that so-called `Gieseker-($\gamma$-) stability' (defined
and discussed below) provides less singular compactifications and is
thus more useful for bundles on higher-dimensional
manifolds~\cite{Friedman}. It is natural to conjecture that this is
the appropriate concept that will emerge in the low-energy geometric
limit of strings.

More precisely, in the limit where string solutions like D-branes are
well approximated by gauge (vector-) bundles supported on cycles of
Calabi--Yau manifolds, one would expect (marginally) $\Pi$-stable
D-branes to emerge as $\gamma$-(semi-) stable bundles.  While this
appears to be true in two dimensions (and therefore automatically in
one dimension where $\gamma$-stability reduces to $\mu$-stability), we
shall see in Sect.~3 that this is not case for the physically relevant
case of three dimensional CY-manifolds.

$\Pi$-stability is believed to be the physically correct condition in
the full quantum theory, and we investigate its consequences in the
low-energy (large volume) limit.  While $\Pi$-stability works on the
triangulated derived category of sheaves, the geometric large volume
version of $\Pi$-stability (here called $\pi$-stability) is applicable
directly to bundles and sheaves.  We show that $\mu$- or
$\gamma$-stability is not enough to capture the spectrum of stable
D6-branes in the large volume limit where the branes may be
represented by bundles.

The purpose of this investigation two-fold: to establish beyond
reasonable doubt that $\pi$-stability is the physically correct
stability concept for CY-filling bundles, and to discover to what it
corresponds in the low-energy algebrao-geometric limit.  Our strategy
is to solve the second problem first by arguing in Sect.~4 for a
specific form of the non-linear deformations of the self-dual
Hermitian Yang--Mills (YM) equations from strings, and then use this
to prove that $\pi$-stability is the correct physical concept by
showing that the solutions of these equations, the non-linear
YM-instantons, exist in the limit of a large enough volume for the CY
iff they are $\pi$-stable.

This result on $\pi$-stability is analogous to the
Donaldson--Uhlenbeck--Yau (DUY) theorem relating solutions of the
Hermitian YM (Hermitian Einstein) equations to $\mu$-stability, and it
reduces to this (as it must) when the string-induced non-linear
deformations are neglected.  Our approach is similar to Leung's
construction~\cite{Leung:1997} of non-linear YM-equations, and we
follow his conventions.  Leung calls these equations the `almost
Hermitian Einstein equations', whose solutions by construction are
$\gamma$-stable bundles.  The deformed YM-equations we consider, on
the other hand, are `canonically' given to us by physics (strings),
and they turn out to be different from Leung's deformations.  Stringy
instantons will therefore not in general be $\gamma$-stable, and our
task is to use these equations to construct the physically correct
topological stability criterion.  This it turns out is precisely the
geometrical limit of $\Pi$-stability, thus closing the circle of ideas
and providing a convincing consistency check on these.

\section{Mathematical ($\mu$- and $\gamma$-) stability}

All stability criteria degenerate to Mumford ($\mu$)-stability when
the base manifold is a curve (one dimension).  By the DUY
theorem~\cite{Donaldson:1985,Uhlenbeck:1986} a vector bundle over an
$n$-dimensional Kähler space is $\mu$-stable\footnote{The general
  $n$-dimensional $\mu$-stability is known as
  Mumford-Takemoto-stability.}, which means that all sub-sheaves $E'
\subset E$ have $\mu_{E'} < \mu_E$, iff the Hermitian YM equation
\begin{equation}
  \omega^{n-1}\wedge F = \frac{\mu_E}{(n-1)!} I
\label{eq:HYM}
\end{equation}
has a unique solution, where $\omega$ is the Kähler form and $I$ is an
identity matrix in colour space.

A vector bundle is \emph{$\mu$-semistable} if all sub-sheaves $E'
\subset E$ have $\mu_{E'} \le \mu_E$.  This definition includes the
stable bundles among the semistable.  A bundle which is semistable but
not stable is called \emph{strictly} semistable.

In higher dimensions Gieseker ($\gamma$)-stability is the commonly
used topological stability condition because it is better behaved.
The $\gamma$-stability condition on a bundle $E$ over a Kähler
manifold $X$ is phrased in terms the \emph{normalised Hilbert
  polynomial} with respect to the Kähler form $\omega$
\cite{Friedman}, which may be calculated using the
Hirzebruch--Riemann--Roch theorem (see {\eg}~\cite{Hartshorne}):
\begin{equation}
  \gamma_E(t)=\frac{1}{\rk E}\int e^{t\omega} \ch(E) \Td(X)
\label{eq:hilb}
\end{equation}
The Chern character is defined by $\ch(E)=\Tr e^F$, where $F$ is the
field strength.  A vector bundle is $\gamma$-stable if all sub-sheaves
$E' \subset E$ have $\gamma_{E'}(t) < \gamma_E(t)$ when $t \to
\infty$.  Using eq.~(\ref{eq:hilb}) one may show that
$\gamma$-stability can be expressed in terms of certain topological
invariants $\muE{k}$ which we will call `generalised slopes'.  They
are defined by
\begin{equation}
\muE{k} := \frac{1}{rk(E)} \int_X \ch_k(E) \frac{\omega^{n-k}}{(n-k)!},
\label{eq:mu-def}
\end{equation}
so that $\mu_E \to \muE{1}$.

The stability condition now takes the form of a series of
inequalities. A bundle is $\gamma$-stable if
\begin{equation}
  \muEp{1} < \muE{1} \text{ for all } E' \subset E,
\label{eq:G1}
\end{equation}
or, if $\muE{1}=\muEp{1}$ for some $E' \subset E$ and there is no $E'
\subset E$ with $\muE{1}>\muEp{1}$,
\begin{equation}
  \muEp{2} < \muE{2} \text{ for all } E' \subset E,
\end{equation}
or, if $\muE{2}=\muEp{2}$ for some $E' \subset E$ and there is no $E'
\subset E$ with $\muE{2}>\muEp{2}$,
\begin{equation}
  \muEp{3} < \muE{3} \text{ for all } E' \subset E,
\end{equation}
and so on.  The first of these inequalities implies $\mu$-stability, a
$\mu$-stable bundle is always $\gamma$-stable.  It also follows that a
$\gamma$-semistable bundle is always $\mu$-semistable, since a
$\gamma$-stable bundle must have $\muEp{1} \le \muE{1}$ for all $E'
\subset E$, but the converse does not hold.

A result analogous to the DUY theorem exists for $\gamma$-stability.
Leung~\cite{Leung:1997} has found a non-linear deformation of the
self-dual YM equation,
\[
  [e^{t\omega + F}\Td(X)]^{(2n)} = \gamma_E(t) I,
\]
which he calls the almost Hermitian Einstein equation, whose solutions
exist iff the instanton bundle is $\gamma$-stable.  The subscript
$(2n)$ means to take the $2n$-form part of the expression.  Leung
showed that when $t \gg 0$ ({\ie} for all $t>T$, where $T$ depends on
the bundle) this equation has a solution iff the bundle is
$\gamma$-stable, which means that $\gamma_{E'}(t) < \gamma_E(t)$ for
all sub-sheaves $E' \subset E$.  This equation is a deformation of the
Hermitian YM equation in the sense that the highest
order term in $t$ is identical to the Hermitian YM equation.
Note that the equation is `perturbative' in the sense that it is only
valid for sufficiently large $t$.  Furthermore, the equation becomes
trivial when taking the trace and integrating.

\section{Physical ($\pi$-) stability}

A quantum field theory with $\mathcal{N}=2$ supersymmetry contains $2$
spinorial symmetry generators $Q_\alpha^A$, $A=1,2$
satisfying
\begin{align}
  \{ Q_\alpha^A, \bar Q_\beta^B \}
&= - 2 \delta^{AB} P_\mu \gamma_{\alpha\beta}^\mu -
2 i \epsilon^{AB} \mathcal{Z} \delta_{\alpha\beta}, \label{eq:Qcomm}
\end{align}
where $\gamma^\mu$ is a Dirac matrix.  $\mathcal{Z}$ is the
\emph{central charge} of the algebra and can be regarded as a complex
number.  The mass $m$ of a state is given by $m \ge |\mathcal{Z}|$,
which is called the Bogomol'nyi-Prasad-Sommerfield (BPS) bound.  If
$m=|\mathcal{Z}|$, the particle is in a `short' (BPS) representation
of the algebra, which preserves half of the supersymmetry of the
theory.  The Hermitian YM equation~(\ref{eq:HYM}) is known to be
related to the BPS condition in $\mathcal{N}=2$ YM
theory~\cite{Harvey:1998gc}.

Douglas et al.~\cite{Douglas:2000ah} have proposed a `stringy
analogue' of $\mu$-stability called $\Pi$-stability.  A brane is
$\Pi$-stable iff, for all sub-branes $\mathcal{E}' \subset
\mathcal{E}$ one has $\varphi_{\mathcal{E}'} < \varphi_{\mathcal{E}}
:= \Arg \mathcal{Z}(\mathcal{E})$.  In the `geometric' or `large
volume limit' ($t\rightarrow\infty$) the central charge is given
by~\cite{Diaconescu:1999vp,Harvey:1998gc}\footnote{The full expression
  for the central charge~\cite{Katz:2002gh,Freed:1999vc} for a D-brane
  on a sub-manifold $S \subset X$ contains the field strength of the
  normal bundle and the A-roof genus, $e^{F_{N_S} / 2}
  \sqrt{\hat{A}(T_S) / \hat{A}(N_S)}$, but this factor reduces to
  $\sqrt{\Td(X)}$ when $S=X$.\label{note:todd}}
\begin{equation}
  \mathcal{Z}(E)\rightarrow Z(E) = - \int_X e^{-i t\omega} \ch(E)\sqrt{\Td(X)},
  \label{eq:ccharge}
\end{equation}
where $\ch(E)$ is the Chern character of the vector bundle $E$, and
$\Td(X)$ is the Todd class of the Calabi--Yau space $X$.  When $X$ is
topologically non-trivial the Todd form mixes the `electric' and
`magnetic' charges of the brane, a phenomenon dubbed the `geometric
Witten effect' in ref.~\cite{Harvey:1998gc}, by analogy with the fact,
discovered by Witten~\cite{Witten:1979ey}, that magnetic monopoles are
dyons on topologically non-trivial spaces.

Using the low-energy expression eq.~(\ref{eq:ccharge}) for the central
charge, one may obtain a hierarchy of inequalities similar to the
conditions derived from $\gamma$-stability.  The calculation is in
this case a bit more involved, and it turns out that the resulting
inequalities depend on the Todd form of the Calabi--Yau, as opposed to
the Gieseker case.  For a torus $T^6$, we get (using the normalisation
$\int \omega^n / n!=1$)
\begin{align*}
  \muEp{1} &< \muE{1}, \\
  \muE{1}\muEp{2} - \muEp{3} &< \muEp{1} \muE{2} - \muE{3}, \\
  \muE{2} \muEp{3} &< \muEp{2} \muE{3}.
\end{align*}
For a quintic hypersurface in $\mathbb{P}^4$, the inequalities
become~\cite{Romelsberger}
\begin{align*}
  \muEp{1} &< \muE{1}, \\
  \muE{1} \muEp{2} - \muEp{3} &< \muEp{1} \muE{2} - \muE{3} \\
 \frac{5}{2}\,\muEp{3}
 + \frac{5}{6}\,\muEp{1} \muE{2}
 + \muE{2} \muEp{3}
 &<
 \frac{5}{2}\,\muE{3}
 + \frac{5}{6}\,\muE{1} \muEp{2}
 + \muEp{2} \muE{3}
\end{align*}
It is clear from these expressions that $\pi$-stability does not
reduce to $\gamma$-stability in the geometric limit, although this has
been speculated in the literature.  Note also that as in the case of
$\gamma$-stability, a $\mu$-stable bundle is always $\pi$-stable, but
a $\pi$-stable bundle is not necessarily $\mu$-stable.  Furthermore, a
$\mu$-semistable bundle does not have to be $\pi$-semistable.  It is
therefore not strictly correct to say that $\pi$-stability always
reduces to $\mu$-stability in the large volume limit.

Leung's almost Hermitian YM equation can therefore not be
the relevant instanton equation for string theory, and
the question arises whether there exists another deformation
of the Hermitian YM equation whose solutions correspond
exactly to $\pi$-stable bundles.  The rest of the paper is
devoted to this problem.

\section{Stringy instantons}

In the geometric limit the conditions for a D-brane (wrapped on a
compact Calabi--Yau manifold) to be BPS may be stated in terms of
geometrical equations.  The BPS condition gives equations both for the
cycle on which the brane is wrapped~\cite{Becker:1995kb} and for the
gauge field strength $F$ and background $B$-field.  The equation for
the gauge field strength was found by Mariño et
al.~\cite{Marino:1999af} for the case of a D6-brane wrapping a Kähler
3-fold with $B=0$ \footnote{Our notation differs slightly in the
  definition of $\theta_E$ from~\cite{Marino:1999af}.  Products are
  exterior products.}
\begin{equation}
      \omega^2 F  - \frac{1}{3}  F^3
  = \cot\theta_E \left( \omega F ^2 - \frac{1}{3} \omega^3\right),
\label{eq:defDUY-3fold}
\end{equation}
where $\theta_E$ is a constant determined by the topology of the gauge
vector bundle $E$.  The value of $\cot\theta_E$ may be found by
integrating eq.~(\ref{eq:defDUY-3fold}).  This equation was recently
confirmed by a world-sheet approach~\cite{Kapustin:2003se}.

Eq.~(\ref{eq:defDUY-3fold}) does not include all instanton
corrections~\cite{Douglas:2000be}, and there has been
speculation~\cite{Minasian:2001na} that the factor $\sqrt{\Td X}$
should be included in this equation.  Our starting point is a
generalisation of the above equation, introducing this
factor\footnote{For a bundle on a proper sub-manifold $S \subset X$,
  the factor $\sqrt{\Td(X)}$ should be replaced as in footnote
  \ref{note:todd}.} and the volume parameter $t$:
\begin{equation}
  \re[e^{-i t\omega + F}\sqrt{\Td(X)}]^{(2n)}
   = \cot\varphi_E(t) \im[e^{-i t\omega + F}\sqrt{\Td(X)}]^{(2n)}
\label{eq:bps-deformed}
\end{equation}
where $\varphi_E(t)$ is the phase of the central charge given by
eq.~(\ref{eq:ccharge}).  By comparing eq.~(\ref{eq:defDUY-3fold}) and
eq.~(\ref{eq:bps-deformed}) one sees that in the case of the torus
$T^6$ where $\Td(X)=1$, $\varphi_E=\theta_E$ (setting $t=1$).  The
condition for $\pi$-stability is therefore in this case equivalent to
$\theta_{E'}<\theta_E$.  We will show that eq.~(\ref{eq:bps-deformed})
has a solution iff the bundle $E$ is $\pi$-stable in the limit $t \gg
0$.  An additional technical assumption we must make~\cite{Leung:1997}
is that the solution of eq.~(\ref{eq:bps-deformed}) is well-behaved in
the limit, in the sense that $\lim_{t\to\infty}|F|/t=0$.  The
necessity of this assumption will become clear below.

Our strategy is to adapt Leung's proof for the case of
$\gamma$-stability to this case.  The main technical tool is geometric
invariant theory (GIT).  We will not here repeat those parts of the
proof which coincide with Leung's case, but summarise the most
important points and emphasise the parts that are different in the two
cases.  We will restrict ourselves to the case when the D-brane is
filling a 3-dimensional Calabi--Yau space, but the theorem should be
straight forward to generalise to other branes.  One should also note
that this procedure is not restricted to eq.~(\ref{eq:bps-deformed}),
but may be used for a rather generic deformation of the Hermitian YM
equations, as long as the stability condition is deformed
correspondingly.

Our first claim is that eq.~(\ref{eq:bps-deformed})
implies that the bundle $E$ is $\pi$-stable.  Let $S$ be any
sub-bundle of $E$.  The key point is to decompose the connection on
$E$ into connections on $S$ and $Q:=E/S$,
\begin{equation}
  A = \begin{pmatrix} A_S & B \\ B^\dagger & A_Q \end{pmatrix},
\end{equation}
so the field strength $F_E$ may be written
\begin{equation}
  F_E = \begin{pmatrix}
      F_S + B B^\dagger &  \partial B \\
       \bar \partial B^\dagger & F_Q + B^\dagger B
  \end{pmatrix}.
\end{equation}

In the case $n=3$, eq.~(\ref{eq:bps-deformed}) is expanded
as
\begin{equation}
  - t^2 \omega^2 F_E + \frac{1}{3} F_E^3 +  F_E \Td_2(X)
   = \cot\varphi_E(t) \left[ \frac{t^3}{3} \omega^3
               - t \omega F_E^2 - t \omega \Td_2(X) \right]
\label{eq:bps-expanded}
\end{equation}
By taking the trace over $S$ only, this becomes
\begin{equation}
\begin{aligned}
\cot\varphi_E(t) &=
      \frac{ \int\Tr_S \left[ - t^2 \omega^2 F_E
                     + F_E^3 / 3 + F_E \Td_2(X) \right] }{
             \int\Tr_S \left[ t^3 \omega^3 / 3
         - t\omega F_E^2 -  t\omega \Td_2(X) \right] } \\
&= \cot\varphi_S(t)
   - \frac{1}{t\rk(S)} \int\Tr_S \omega^2 BB^\dagger
   + \Delta,
 \end{aligned}
\label{eq:varphi}
\end{equation}
\[
\begin{aligned}
 \Delta &=
   \frac{ \int\Tr_S \left[ (F_E^3 - F_S^3) / 3
                           + BB^\dagger \Td_2(X) \right] }{
            \int\Tr_S \left[ t^3 \omega^3 / 3
                     -  t\omega F_E^2 - t\omega \Td_2(X) \right] }
 \\  &\phantom{={}} 
    + \frac{\int\Tr_S \omega^2 BB^\dagger }{t\rk(S) }
      \frac{\int\Tr_S \left[- t\omega F_E^2 - t\omega \Td_2(X) \right] }{
            \int\Tr_S \left[ t^3 \omega^3 / 3
                     -  t\omega F_E^2 - t\omega \Td_2(X) \right]}
 \\  &\phantom{={}} 
 - \frac{ \int\Tr_S \left[ -  t^2 \omega^2 F_S
                             + F_S^3 / 3 + F_S \Td_2(X) \right]
          \int\Tr_S \left[ t\omega ( F_S^2 - F_E^2 ) \right] }{
             \int\Tr_S \left[ t^3 \omega^3 / 3
                     -  t\omega F_E^2 - t\omega \Td_2(X) \right]
             \int\Tr_S \left[ t^3 \omega^3 / 3
                     -  t\omega F_S^2 - t\omega \Td_2(X) \right] }
\end{aligned}
\]
The term $\int \Tr_S \omega^2BB^\dagger$ is positive, see
{\eg}~\cite[p.~79]{Griffiths-Harris}.  It is therefore enough to show
that $|\Delta| < |\int\Tr_S \omega^2 BB^\dagger| / t\rk(S)$.  Leung
shows that if $\mathcal{T}_k$ is a term containing a product of $k+1$
or fewer terms of the type $B^\dagger B$, $\bar\partial B$, $\partial
B^\dagger$, $F_S$ and $F_Q$, then the assumption
$\lim_{t\to\infty}|F|/t=0$ leads to the conclusion that $\mathcal{T}_k
/ t^k \ll \int |B|^2$.

We can choose a $t$ large enough so that the denominators of the
fractions in $\Delta$ are arbitrarily close to $t^3\rk(S)$, and by
Leung's arguments, each of these terms is then not greater than a
small fraction of $\int |B|^2$.  Thus, since we get $\cot\varphi_E(t)
- \cot\varphi_S(t) < 0$ for all sub-bundles $S \subset E$, as
$t\to\infty$ we have $\varphi_{S}(t) < \varphi_E(t)$, which proves the
first part of the proposition.

For the converse we must show that there exists a solution of
eq.~(\ref{eq:bps-expanded}) for any $\pi$-stable bundle $E$, so assume
now that the bundle $E$ is $\pi$-stable.  We expand the topological
parameter $\cot\varphi_E(t)$ as a series in $1/t$.  Defining
$\pi_E(t)=-t\cot\varphi_E(t)$, we may write
eq.~(\ref{eq:bps-expanded}) as
\begin{multline}
   \omega^2 F
   +  t^{-2} \left( \mu_1 \omega F^2
                 - \frac{1}{3} F^3
                 + \mu_1 \omega \Td_2(X) \right) \\
   + t^{-4} \left( \piE{2} \omega  F^2
                 + \piE{2} \omega  \Td_2(X) \right)
   + \mathcal{O}(t^{-6})
   = \pi_E(t) \frac{\omega^3}{3},
\label{eq:defDUY}
\end{multline}
where
\begin{equation}
\begin{aligned}
  \pi_E(t)
  &= \piE{1} + t^{-2} \piE{2} + t^{-4} \piE{3} + \mathcal{O}(t^{-6}) \\
  &= \muE{1} + t^{-2} \left( \muE{1} \muE{2} - \muE{3}  + \muE{1} \tau \right)\\
  & \qquad\qquad\qquad\qquad+
       t^{-4} \muE{2} ( \muE{1} \muE{2} - \muE{3} + \muE{1} \tau ) 
   + \mathcal{O}(t^{-6}),
\end{aligned}
\end{equation}
and $\tau=\int\omega\Td_2(X)$.  For $0<\varphi_E<\pi$, $\pi$-stability
is now equivalent to $\pi_{E'}(t) < \pi_E(t)$, and we must show that
there is a solution to eq.~(\ref{eq:defDUY}) in this case.

An important property of (strictly) $\mu$-semistable bundles is that
they may be `deconstructed' into a sequence (or \emph{filtration}) of
\emph{$\mu$-stable} bundles.  This is the Jordan-Hölder theorem:
\emph{If $E$ is a $\mu$-semistable vector bundle, there is a
  filtration $E \supset E_1 \supset \cdots \supset E_k \supset 0$ such
  that each $Q_j := E_j / E_{j+1}$ is $\mu$-stable and $\mu_1(Q_j) =
  \mu_1(E)$ for each $j$.}

Since the quotients $Q_j$ are $\mu$-stable, there exist connections
$A^{(j)}$ on each of them satisfying the Hermitian YM equation
$\omega^{n_j-1} F^{(j)}=\muE{1}\omega^n_j/n_j!$.
Therefore, there exists a (Hermitian) connection $A$ on $E$,
\begin{equation}
  A = \begin{pmatrix} A^{(k)} & B^{(k-1,k)} & \cdots & B^{(1,k)} \\
                      -B^{(k-1,k)\dagger} & A^{(k-1)} & \cdots & B^{(1,k-1)} \\
                      \vdots & \vdots & \ddots & \vdots \\
                      -B^{(1,k)\dagger} & -B^{(2,k)\dagger} & \cdots & A^{(k)}
      \end{pmatrix}.
\label{eq:JH-DUY}
\end{equation}
The off-diagonal parts $B^{(j,\ell)}$ are uniquely
given~\cite{Leung:1997} up to a scalar factor from the extension
$0 \to E_{j+1} \to E_j \to Q_j \to 0$.
$B^{(j,\ell)}$ is the component in $\Hom(Q_j,Q_\ell)$ of the
representative of the extension class of this extension.

The Jordan-Hölder theorem is applicable to our case since a
$\pi$-stable bundle is always $\mu$-semistable.  Assume now that a
bundle has a connection satisfying eq.~(\ref{eq:bps-deformed}).  Then
$E$ is $\mu$-semistable.  This may be proven by noting that there
exists a metric on $E$ for all positive constants $\delta$ such that
its curvature satisfies $| \wedge R - \mu_1 I | < \delta$, where
$\wedge R = *(\omega^{n-1} R)$.  This follows since there is a
connection satisfying eq.~(\ref{eq:bps-deformed}) for any $t$
sufficiently large.  Since $E$ is $\mu$-semistable, we may then use
the Jordan-Hölder theorem to create a filtration $E \supset E_1
\supset \cdots \supset E_k \supset 0$.

Following Leung we use the simplest case, in which the bundle has a
short Jordan-Hölder filtration, as an example: $E \supset S \supset 0$.
We define $Q := E/S$.  By the Jordan-Hölder theorem we then have
$\mu_Q^{(1)}=\mu_S^{(1)}=\mu_E^{(1)}$.  We will also assume for now
that $E$ is $\pi$-stable ``at second order'', {\ie} that
$\piEp{1}=\piE{1}$ and $\piEp{2}<\piE{2}$ for all sub-bundles $E'$.

Since we want to find a solution to eq.~(\ref{eq:defDUY}) which is a
perturbation of the DUY equation, we now perturb the connection
(\ref{eq:JH-DUY}) with corrections proportional to $t^{-1}$ and
$t^{-2}$, and let
\begin{equation}
A_E = \begin{pmatrix}
 A_S + (h_S^{-1}) \partial h_S / t^2 &
 B/t - \bar\partial (\phi^\dagger) / t^2\\
 - B^\dagger / t + \partial\phi / t^2&
 A_Q + (h_Q^{-1}) \partial h_Q / t^2
\end{pmatrix},
\end{equation}
where we will determine the proper normalisation of $B$ later.
Eq.~(\ref{eq:defDUY}) now becomes
\begin{equation}
  \left( \bar\partial ( h_S^{-1} \partial h_S )
                       - B B^\dagger \right)  \omega^2
   = \piE{2} \frac{\omega^3}{3}
      - \left[ \mu_1 \omega F_{0S}^2 - \frac{1}{3} F_{0S}^3
               + \mu_1 \omega \Td_2(X) \right]
           + \mathcal{O}(\frac{1}{t})
\label{eq:defDUY1}
\end{equation}
\begin{align}
  \left( \bar\partial ( h_Q^{-1} \partial h_Q )
                       - B^\dagger B \right) \omega^2
   &= \piE{2} \frac{\omega^3}{3}
      - \left[ \mu_1 \omega F_{0Q}^2 - \frac{1}{3} F_{0Q}^3
               + \mu_1 \omega \Td_2(X) \right]
           + \mathcal{O}(\frac{1}{t})\\
   \bar\partial \partial \phi \,\, \omega^2
    &= \mathcal{O}(\frac{1}{t})
\end{align}
The following theorem~\cite{Uhlenbeck:1986,Leung:1997} tells us when
these equations have solutions: \emph{Let $E$ be a $\mu$-stable bundle
  over a compact Kähler manifold $X$.  Suppose $H_0$ is an
  endomorphism of $E$ such that $\int \Tr_E H_0 = 0$.  Then there
  exists a unique positive self-adjoint endomorphism $h$ of $E$ with
  $\det h=1$, which solves $\bar\partial ( h^{-1} \partial h ) =
  H_0.$}

This is where that the requirement for $\pi$-stability enters.  In
order to fulfil the requirement for the theorem, we must have
\begin{equation}
  0 \le \int |B|^2 = (\piE{2} - \piS{2}) \rk(S).
\label{eq:B}
\end{equation}
Eq.~(\ref{eq:B}) can only be satisfied when $\piS{2} \le \piE{2}$.
When this is the case, we have $\int \Tr_S H_0 = 0$.  Therefore,
eq.~(\ref{eq:defDUY1}) with $t\to\infty$ always has a solution when
$E$ is $\pi$-stable.

This can be extended to $t<\infty$ by rescaling $B$ by a parameter
depending on $h_S$, $h_Q$ and $\phi$.  One can show that there is a
rescaling parameter such that $\int \Tr_S H = 0$, where $H$ is given
by eq.~(\ref{eq:defDUY-3fold}), thus the above theorem also holds for
$t<\infty$.

Following Leung, this example may be generalised to
\[
  \piS{n} = \piE{n} \quad (n=1,2,\ldots,m)\qquad\qquad
  \piS{m+1} < \piE{m+1}.
\]
Using the above approach order by order in $t$, there will exist a connection
\begin{equation}
  A^{(n)} = \begin{pmatrix}
   A^{(n-1)}_S + h_{(n)S}^{-1} \partial h_S / t^{2n} &
  B/t^m -  \bar\partial \phi^\dagger - \delta_n^\dagger / t^2\\
  B^\dagger / t^m - \partial \phi - \delta_n / t^2 &
  A^{(n-1)}_Q + h_{(n)S}^{-1} \partial h_S  / t^{2n}
\end{pmatrix}
\end{equation}
where $\delta_n = \partial \phi_{\frac{m}{2}+1} / t^{m+2} + \cdots +
\partial \phi_n/ t^{2n}$ for $n>\frac{m}{2}$, and $\delta_n=0$
otherwise.  This connection will solve eq.~(\ref{eq:defDUY-3fold}) to
order $t^{-2m}$.  For higher orders, one must again rescale $B$.

Finally, assume that $E$ is a general $\pi$-stable bundle, which then has
a Jordan-Hölder filtration
$E = E_0 \supset E_1 \supset \cdots \supset E_k \supset E_{k+1} = 0$,
and the bundles $Q_j := E_j / E_{j+1}$ are  $\mu$-stable and have the
same slope $\muE{1}$ as $E$.  The connection on $E$ will now be
\[
  A = \begin{pmatrix} A^{(k,k)} & A^{(k,k-1)} & \cdots & A^{(k,1)} \\
A^{(k-1,k)} & A^{(k-1,k-1)} & \cdots & A^{(k-1,1)} \\
\vdots & \vdots & \ddots & \vdots \\
A^{(1,k)} & A^{(1,k-1)} & \cdots & A^{(1,1)}
\end{pmatrix},
\]
with
\[
A^{(j,j)}_\mu = A_{0\mu}^{(j)} + \frac{1}{t^2}  h_j^{-1} \partial h_j
\qquad\qquad
  A^{(i,j)}_\mu = \frac{1}{t} B_{(i,j)} - \frac{1}{t^2} \bar\partial \phi^\dagger_{(i,j)}  \qquad i>j 
\]
\[
A^{(i,j)}_\mu = -\frac{1}{t} B_{(j,i)}^\dagger  + \frac{1}{t^2} \partial \phi_{(j,i)}  \qquad i<j
\]
with one $B_{(i,j)}$ and one $\phi_{(i,j)}$ for each pair
$(i,j),\,i>j$.  The above procedure can then be repeated in this
general case, and we may finally conclude that
eq.~(\ref{eq:bps-deformed}) has a solution iff the bundle $E$ is
$\pi$-stable.

\section{Conclusion and outlook}

We have related $\pi$-stability of a D-brane in the large volume limit
to a geometric condition on the field strength of the gauge bundle,
eq.~(\ref{eq:bps-deformed}), which is a deformation of the well-known
Hermitian YM equation.  $\Pi$-stability is the correct stability
condition in the full (quantum corrected) theory and we propose
eq.~(\ref{eq:bps-deformed}) to be the correct instanton equation for
describing BPS D-branes geometrically (in the large volume limit),
generalising the Hermitian YM equation which describes BPS instantons
in (undeformed) YM theory.

The string-deformed low-energy YM field equations are just the
fossilised low-energy remains of a huge framework of great beauty,
consistency and subtlety whose low-energy implications have in the
past had significant impact on algebraic geometry ({\eg} mirror
symmetry).  It is therefore not unreasonable to hope that geometric
$\pi$-stability will be of use as well.  While $\gamma$-stability is a
refinement of $\mu$-stability, a `finer sieve' used to select
semi-stable bundles, $\pi$-stability is quite distinct from
$\gamma$-stability in three dimensions, and may therefore be expected
to give rise to genuinely new and interesting compactifications of the
moduli spaces studied in algebraic geometry.

The large volume limit has in this paper been taken with $B=0$, which
is the traditional `geometric' limit.  However, it would be
interesting to see the consequences of including also a $B$-field in
the geometric picture.  Shifting the $B$-field by an integral form is
equivalent to tensoring the bundle with a line bundle, which in the
case of $\gamma$-stability does not affect stability.  For
$\pi$-stability however, an integral shift of the $B$-field may affect
stability, as found in~\cite{Douglas:2000qw}.  This means that
including the $B$-field may extend the moduli space of $\pi$-stable
bundles in a non-trivial way.

\section{Acknowledgements}

We are grateful to M. Douglas, W. Lerche and G. Ellingsrud for helpful
discussions, and especially to C. R\"omelsberger for patiently
explaining $\Pi$-stability and for collaboration at an early stage of
this project.  This work was supported by the Research Council of
Norway and was made possible by the generosity of the CERN Theory
Group.

\providecommand{\href}[2]{#2}\begingroup\raggedright\endgroup

\end{document}